# Variational Approach to Solving the Spectral Boltzmann Transport Equation in Transient Thermal Grating for Thin Films

Vazrik Chiloyan[a], Lingping Zeng[a], Samuel Huberman[a], Alexei A. Maznev[b], Keith A. Nelson[b], Gang Chen[a]*

[a]Department of Mechanical Engineering, Massachusetts Institute of Technology, Cambridge, Massachusetts 02139, USA
[b]Department of Chemistry, Massachusetts Institute of Technology, Cambridge, Massachusetts 02139, USA

**Abstract**

The phonon Boltzmann transport equation (BTE) is widely utilized to study non-diffusive thermal transport. We find a solution of the BTE in the thin film transient thermal grating (TTG) experimental geometry by using a recently developed variational approach with a trial solution supplied by the Fourier heat conduction equation. We obtain an analytical expression for the thermal decay rate that shows excellent agreement with Monte Carlo simulations. We also obtain a closed form expression for the effective thermal conductivity that demonstrates the full material property and heat transfer geometry dependence, and recovers the limits of the one-dimensional TTG expression for very thick films and the Fuchs-Sondheimer expression for very large grating spacings. The results demonstrate the utility of the variational technique for analyzing non-diffusive phonon-mediated heat transport for nanostructures in multi-dimensional transport geometries, and will assist the probing of the mean free path (MFP) distribution of materials via transient grating experiments.

*Corresponding author: gchen2@mit.edu

# I. INTRODUCTION

Recent years have witnessed intensifying research of phonon-mediated thermal transport on the micro- and nano-scale[1–13] stimulated by diverse technology drivers such as thermal management of microelectronic devices[14,15] and nanostructured thermoelectric materials[16–18]. It has been realized that the heat diffusion equation, which has been the workhorse of thermal transport science and engineering, becomes inadequate at distances comparable to the mean free path (MFP) of heat-carrying phonons. Recently, non-diffusive heat conduction has been observed experimentally with optical-based techniques such as time-domain thermoreflectance[4,5,19–23] (TDTR), frequency-domain thermoreflectance[2,24] (FDTR), and transient thermal grating[10–12,25] (TTG). These experiments typically report effective thermal conductivities, extracted by applying the Fourier heat conduction equation to the experimental configuration, that are lower than those of the bulk materials. These effective thermal conductivities vary as a function of experimental variables such as the laser heating modulation frequency[1,2], the heat source diameter[23], the grating spacing[10,11], etc. Theoretical analysis of non-diffusive transport is typically performed using the Boltzmann transport equation (BTE), with the frequency dependent mean free path (MFP) distribution as an input to the equation. Inversely, one can use these experimental data to reconstruct the MFP distribution[12,23] in the form of a thermal conductivity accumulation function using a suppression function obtained through solving the BTE. Unlike the Fourier heat conduction equation, however, the BTE is notoriously difficult to solve, especially when heat conduction involves multiple dimensions. Numerical techniques for direct solution of the BTE or via Monte Carlo simulation have been developed[1,5–7,9,25–30], but they are difficult to apply to existing experimental geometries. We recently developed a variational approach to solve the BTE using the temperature distribution obtained from the Fourier heat conduction as the trial function and demonstrated the effectiveness of this method by applying it to the one-dimensional thermal grating relaxation in bulk material[3]. Here, we demonstrate the utility of this approach for analyzing heat transport for nanostructures in multi-dimensional geometries and in the presence of phonon scattering at boundaries, which is more relevant to realistic experimental conditions. Specifically, we analyze the relaxation of a thermal grating in a suspended thin membrane[11]. The consideration of size effects from the thin film is critical to explain the experimentally obtained thermal decay profiles, especially for very thin films[12] and can reveal the suppression of thermal conductivity and deviation from the Fourier heat conduction solution both from the size effects of the grating spacing and the film thickness. This is a nontrivial extension since the spectral BTE in a multi-dimensional geometry is notoriously difficult to solve.

In the TTG experiment, two pulsed laser beams are crossed in order to generate a sinusoidal heating profile of spatial period $\lambda$. As the thermal grating relaxes, the diffraction of a probe laser from the sample captures the thermal decay dynamics. The TTG technique has been used to observe non-diffusive thermal transport in silicon membranes[11,12] and gallium arsenide[10] at room temperature, and can enable the extraction of the MFP distribution[23]. The simplicity of the experimental geometry (including, above all, the absence of interfaces between different materials) makes it an important system to study theoretically. The TTG geometry has been well studied in the one-dimensional bulk limit of a very large membrane thickness. In this case, the transport is purely one-dimensional and occurs only in the in-plane direction of the film between hot and cold spots generated by the laser intensity profile. Various approaches to solving this case include the two-fluid framework[25] with simplifying assumptions about the scattering of high and low frequency phonons, an exact numerical solution with finite differences[13], a temporal

Fourier transform approach for which the temperature in the frequency domain was obtained analytically[9]. More recently, we developed a variational method[3] with which the thermal decay rate, effective thermal conductivity, and corresponding suppression function were calculated analytically. This closed form expression was the first to show the full material property dependence of the suppression function and addresses the non-universality of the suppression function.

The BTE is considerably harder to solve in a multi-dimensional geometry[4,5] and in the presence of boundary scattering. The thin film TTG geometry has been studied previously with the Monte Carlo approach[6]. In this work, we demonstrate the ability of the variational approach to provide an analytical form for the temperature profile and the associated thermal decay rate. This provides the ability to study the full material and geometry dependence of the thermal transport in the TTG configuration. We demonstrate that results from this variational approach are in excellent agreement with Monte Carlo simulation on Si and PbSe. The variational approach offers the ability to accurately solve the BTE and study a wide range of materials in the non-diffusive regime with a closed form expression, and computationally a much faster way to study non-diffusive transport over a broad range of length scales for the film thickness and the thermal grating period.

## II. THE SPECTRAL BTE AND TEMPERATURE EQUATION

We begin with the spectral Boltzmann transport equation under the relaxation time approximation for the spectral energy density[13,31,32]

$$\frac{\partial g_\omega}{\partial t} + \vec{v}_\omega \cdot \vec{\nabla} g_\omega = \frac{g_0 - g_\omega}{\tau_\omega} \tag{1}$$

where $g_\omega$ is the phonon energy density per unit frequency interval per unit solid angle above the reference background energy, related to the distribution function as $g_\omega = \frac{\hbar\omega D(\omega)}{4\pi}(f_\omega - f_0(T_0))$. $v_\omega$ is the group velocity, $\tau_\omega$ is the relaxation time, and $g_0$ is the equilibrium energy density, given by $g_0 \approx \frac{1}{4\pi}C_\omega(T-T_0)$ in the linear response regime. In the TTG experiment, the temperature initially has a sinusoidal periodic profile as $T(x,t=0) = T_0 + T_m e^{iqx}$ in complex form where $T_m$ is the initial amplitude of the spatial variation and $q = 2\pi/\lambda$ is the wavevector for the grating sinusoidal profile. We assume that the initial temperature profile is uniform across the thickness of the film (depicted in Fig. 1) as is often the case in experiment[25]. We require $T_m \ll T_0$ to ensure the linear response regime. We expect the temperature profile to obey $T(x,z,t) = T_0 + T_m e^{iqx} h(z,t)$ where $h(z,t)$ is the non-dimensional temperature that satisfies $0 \leq h(z,t) \leq 1$, $h(z,t=0) = 1$ and $h(z,t \to \infty) = 0$.

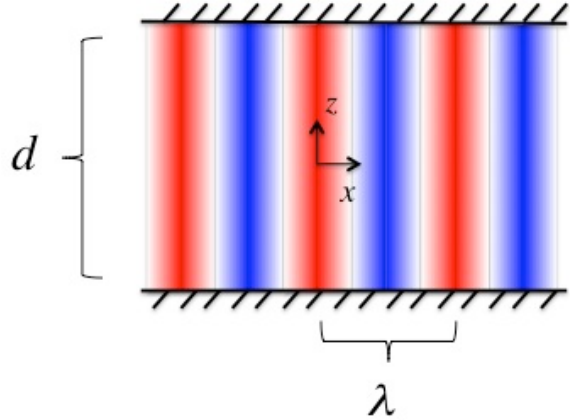

FIG 1: Schematic showing the geometry and coordinate system for the thin film. The thickness of the film $d$ is depicted separating the adiabatic top and bottom surfaces. The grating period $\lambda$ shows the separation

between hot spots (red) and cold spots (blue), which are uniform vertically across the membrane thickness initially ($t = 0$).

Similarly, the distribution function will be given by $g_\omega = e^{iqx}\tilde{g}_\omega$ and the equilibrium distribution by $\tilde{g}_0 = \frac{C_\omega T_m}{4\pi}h(z,t)$. The BTE now takes the form:

$$\frac{\partial \tilde{g}_\omega}{\partial t} + v_\omega\left(iq\mu_x\tilde{g}_\omega + \mu_z\frac{\partial \tilde{g}_\omega}{\partial z}\right) = \frac{\tilde{g}_0 - \tilde{g}_\omega}{\tau_\omega} \qquad (2)$$

where we have removed the y-dependence due to the translational symmetry, and define $\mu_x, \mu_z$ as the direction cosines in the $x$ (in-plane) and $z$ (cross-plane) directions, respectively. If one measures a polar angle $\theta$ and azimuthal angle $\phi$ from the $z$-axis, then in that case the direction cosines are simply $\mu_z = \cos(\theta)$ and $\mu_x = \sin(\theta)\cos(\phi)$, but we will keep the form general. We utilize the Laplace transform in time and the initial condition of the temperature profile to reduce this partial differential equation to an ordinary differential equation:

$$\frac{\partial \hat{\tilde{g}}_\omega}{\partial u} + \frac{1+s\tau_\omega+i\eta_\omega\mu_x}{2\text{Kn}_\omega\mu_z}\hat{\tilde{g}}_\omega = \frac{\hat{\tilde{g}}_0 + \tau_\omega\frac{C_\omega T_m}{4\pi}}{2\text{Kn}_\omega\mu_z} \qquad (3)$$

where we have replaced $z$ by a non-dimensional variable $u = 2z/d$ where $d$ is the thickness of the thin film so that $-1 < u < 1$. Furthermore, we have defined the following non-dimensional quantities utilizing the mean free path $\Lambda_\omega = v_\omega\tau_\omega$ to characterize the size effect of the grating spacing, $\eta_\omega = q\Lambda_\omega$, and the film thickness, $\text{Kn}_\omega = \Lambda_\omega/d$, given by the Knudsen number. To simplify the notation, we will utilize the term $V \equiv \frac{1+s\tau_\omega+i\eta_\omega\mu_x}{2\text{Kn}_\omega\mu_z}$ to group the variables in a compact form. We impose adiabatic, diffuse boundary conditions at the top and bottom of the thin film,

$$\hat{\tilde{g}}_\omega(u=-1,s,\mu_x,\mu_z>0)=\sigma_1$$
$$\hat{\tilde{g}}_\omega(u=1,s,\mu_x,\mu_z<0)=\sigma_2 \qquad (4)$$

where we define the following solid angle integrals:

$$\sigma_1 \equiv \frac{1}{\pi}\int d\Omega\,\Theta(\mu_z)\mu_z\hat{\tilde{g}}_\omega(u=-1,s,\mu_x,-\mu_z)$$
$$\sigma_2 \equiv \frac{1}{\pi}\int d\Omega\,\Theta(\mu_z)\mu_z\hat{\tilde{g}}_\omega(u=1,s,\mu_x,\mu_z) \qquad (5)$$

which are proportional to the spectral energy flux approaching the bottom and top walls of the thin films, respectively. Note the integrations are only over hemispheres representing flux towards the corresponding surface and not the entire solid angle. We have utilized the Heaviside step functions $\Theta(x)$ to restrict the integration region appropriately.

Solving Eq. (3) and applying the boundary conditions of Eq. (4) yields:

$$\hat{\tilde{g}}_\omega(u,s,\mu_x,\mu_z)=\Theta(\mu_z)\left\{\sigma_1\exp(-V(u+1))+\int_{-1}^u du'\exp(-V(u-u'))\frac{\hat{\tilde{g}}_0(u',s)+\tau_\omega\frac{C_\omega T_m}{4\pi}}{2\text{Kn}_\omega\mu_z}\right\}$$
$$+\Theta(-\mu_z)\left\{\sigma_2\exp(-V(u-1))+\int_1^u du'\exp(-V(u-u'))\frac{\hat{\tilde{g}}_0(u',s)+\tau_\omega\frac{C_\omega T_m}{4\pi}}{2\text{Kn}_\omega\mu_z}\right\} \qquad (6)$$

where the first term describes phonons flowing towards the positive z-direction (top boundary) and the second term describes phonons flowing towards the negative z-direction (bottom boundary). Integrating the solution of Eq. (6) over the solid angle hemispheres at the walls from Eq. (5) yields coupled equations for the energy approaching the walls of the thin film, $\sigma_1,\sigma_2$.

Solving this coupled equation, and utilizing the even symmetry of the system about the center line $u = 0$ of the thin film yields:

$$\sigma_1 = \sigma_2 = \frac{2}{1-2F_3(2)} \int_{-1}^{1} du' \frac{\hat{\tilde{g}}_0(u',s) + \tau_\omega \frac{C_\omega T_m}{4\pi}}{2\text{Kn}_\omega} F_2(1+u') \tag{7}$$

where we have defined the following integral function for convenience:

$$F_n(u) \equiv \frac{1}{2\pi} \int d\Omega \Theta(\mu_z) \mu_z^{n-2} \exp(-Vu) \tag{8}$$

The symmetry of the coordinate system we have chosen requires that the temperature, and thus the equilibrium energy density, be even in the spatial variable $u$. Thus this completes the solution for the spectral energy density in terms of the equilibrium energy density (and thus the temperature) when combining the results of Eqs. (6-8).

Integrating Eq. (1) with respect to the solid angle and all phonon modes yields the equilibrium condition in the spectral case[31]. The equilibrium condition in this case can be expressed as:

$$2\int d\omega \frac{1}{\tau_\omega} \hat{\tilde{g}}_0(u,s) = \int d\omega \frac{1}{\tau_\omega} \frac{1}{2\pi} \int d\Omega \hat{\tilde{g}}_\omega(u,s,\mu_x,\mu_z) \tag{9}$$

Note that the integral over $\omega$ is a compact notation we use for simplicity where it implies a sum over all phonon polarizations and corresponding frequencies of those branches[33]. Inputting the solution of Eq. (6) into Eq. (9), and inputting the expression for the non-dimensional temperature $\hat{\tilde{g}}_0(u,s) = \frac{C_\omega T_m}{4\pi} \hat{h}(u,s)$, we obtain the integral equation governing the temperature distribution:

$$\hat{h}(u,s) \int d\omega \frac{C_\omega}{\tau_\omega} = \int d\omega \frac{C_\omega}{\tau_\omega} \int_{-1}^{1} du' \frac{\hat{h}(u',s) + \tau_\omega}{4\text{Kn}_\omega} \left\{ F_1(|u-u'|) + \frac{2F_2(1+u')}{1-2F_3(2)} [F_2(1+u) + F_2(1-u)] \right\} \tag{10}$$

Notice that this is an integral equation in the spatial variable *u*, which, after solving, would require an inverse Laplace transform to obtain the temporal decay of the temperature profile.

## III. VARIATIONAL APPROACH

We seek an approximate solution of Eq. (10) by using a trial function obtained using the heat diffusion equation. In the thin film TTG geometry, the latter yields a very simple exponentially decaying solution of $T(\vec{r},t) = T_0 + T_m e^{iqx} e^{-\alpha q^2 t}$, where $\alpha$ is the thermal diffusivity. Following the approach previously utilized for the one-dimensional TTG[3], we take a trial solution $\bar{h}(u,t) = \exp(-\gamma t)$ where the decay rate $\gamma = q^2 \alpha_{eff}$ is determined by the "effective" thermal diffusivity $\alpha_{eff}$. By treating the latter as a parameter, we seek to optimize the chosen simple trial function to get the best approximate solution of Eq. (10).

As was demonstrated previously[3], one can take both a mathematical approach, seeking to reduce the least squares error of the error residual of the temperature equation from Eq. (10), or one can impose physical constraints that the trial function should satisfy. Since our trial function has only one variational parameter, it would suffice to impose a single condition to optimize the solution. The physical condition we pick is to demand energy conservation to hold over the thin film control volume considering the whole time decay. In this geometry, we take a control volume over the thickness of the film, and for convenience of a width equal to half of the grating period, centered at a peak in the temperature, where the temperature is above the average background in order to observe heat flux outwards laterally towards the troughs in the temperature spatial profile. By considering the entire decay time, energy conservation demands that the total energy initially deposited by the heating laser pulse into the control volume must be equal to the total energy that flows out of the control volume over the entire decay time. As the top and bottom surfaces of the thin film are adiabatic, we need not consider the flux in the $z$ (cross plane) direction, $q_z$, and thus energy only flows out due to in-plane flux out of the control volume. Integrating over all time, and over the thickness of the thin film in the $z$ direction, and over the width of the control volume in the $x$ direction yields in complex form:

$$CT_m \frac{\lambda d}{\pi} = 2i \int_{-\frac{d}{2}}^{\frac{d}{2}} dz \int_0^\infty dt \, \tilde{q}_x(z,t) \qquad (11)$$

where $q_x$ is the heat flux in the $x$ (in-plane) direction, given by $q_x(x,z,t) = e^{iqx}\tilde{q}_x(z,t)$ due to the periodicity of the heating profile. Eq. (11) simply says that the total initial energy must be equal to the total flux away in the in-plane direction from peak to trough integrated over time.

Beyond a physical demand for energy conservation, Eq. (11) has also a mathematical benefit. In the 1D TTG model, it can be shown that imposing this physical constraint as demonstrated previously[3], makes it such that the area under the temperature decay curve from the trial exponential solution matches the area under the exact temperature decay. While this cannot be strictly proven in the thin film TTG problem due to the $z$ dependence in the problem, the benefit of this is that a function that starts at unity will be constrained to match the area of the actual decay and for the monotonically decaying temperature profile, this will yield an excellent approximation, as will be seen in Fig. 2.

The heat flux is obtained by integrating the spectral energy density over the solid angle and phonon frequencies in the form $\hat{\tilde{q}}_x(u,s) = \int d\omega v_\omega \int d\Omega \mu_x \hat{\tilde{g}}_\omega(u,s,\mu_x,\mu_z)$. Inputting the spectral energy density expression of Eq. (6) and integrating, we obtain:

$$\hat{\tilde{q}}_x(u,s) = T_m \int d\omega C_\omega v_\omega \int_{-1}^1 du' \frac{\hat{h}(u',s) + \tau_\omega}{4\mathrm{Kn}_\omega} \left\{ G_1(|u-u'|) + \frac{2F_2(1+u')}{1-2F_3(2)}[G_2(1+u) + G_2(1-u)] \right\} \qquad (12)$$

where the following solid angle functions have been introduced for convenience:

$$F_n(u) = \frac{1}{2\pi} \int d\Omega \Theta(\mu_z) \mu_z^{n-2} e^{-Vu}$$
$$G_n(u) = \frac{1}{2\pi} \int d\Omega \Theta(\mu_z) \mu_x \mu_z^{n-2} e^{-Vu}$$
(13)

The in-plane heat flux can be integrated over the thickness of the film and after inputting the trial exponential function, whose Laplace transform is given by $\hat{h}(u,s) = \frac{1}{s+\gamma}$, and taking $s$ to be zero to integrate over all time, the conservation equation of Eq. (11) can be solved to yield the thermal decay rate:

$$\gamma = \frac{\int d\omega \frac{C_\omega}{\tau_\omega} \left\{ 1 - \frac{1}{\eta_\omega} \arctan(\eta_\omega) + \Psi(\eta_\omega, \mathrm{Kn}_\omega) \right\}}{\int d\omega C_\omega \left\{ \frac{1}{\eta_\omega} \arctan(\eta_\omega) - \Psi(\eta_\omega, \mathrm{Kn}_\omega) \right\}} \qquad (14)$$

where we have defined the following functions given by solid angle integrals:

$$\Psi(\eta_\omega, \mathrm{Kn}_\omega) \equiv \psi_2 - \frac{\psi_1^2}{\psi_0}$$
$$\psi_n(\eta_\omega, \mathrm{Kn}_\omega) \equiv \frac{1}{2\pi} \int d\Omega \Theta(\mu_z) \frac{\mathrm{Kn}_\omega \mu_z}{(1+i\eta_\omega \mu_x)^n} \left[ 1 - \exp\left(-\frac{1+i\eta_\omega \mu_x}{\mathrm{Kn}_\omega \mu_z}\right) \right]$$
$$= \int_0^1 d\phi \int_0^1 d\mu \frac{\mathrm{Kn}_\omega \mu}{\left(1+i\eta_\omega \sqrt{1-\mu^2} \cos(2\pi\phi)\right)^n} \left[ 1 - \exp\left(-\frac{1+i\eta_\omega \sqrt{1-\mu^2} \cos(2\pi\phi)}{\mathrm{Kn}_\omega \mu}\right) \right]$$
(15)

which can be viewed as generalizations of the exponential integral functions[34].

Using the relation of the thermal conductivity to the thermal decay rate from the Fourier heat conduction solution $k = \gamma C / q^2$, we obtain the effective thermal conductivity for the thin film TTG:

$$k = \frac{\frac{1}{3} \int d\omega C_\omega v_\omega \Lambda_\omega \frac{3}{\eta_\omega^2} \left\{ 1 - \frac{1}{\eta_\omega} \arctan(\eta_\omega) + \Psi(\eta_\omega, \mathrm{Kn}_\omega) \right\}}{\frac{1}{C} \int d\omega C_\omega \left\{ \frac{1}{\eta_\omega} \arctan(\eta_\omega) - \Psi(\eta_\omega, \mathrm{Kn}_\omega) \right\}} \qquad (16)$$

Note the complex material property dependence of the effective thermal conductivity, in that we have two integrals over phonon properties appearing. We take the limits of large membrane thickness and large grating spacing and verify the expected limits of the one-dimensional TTG expression[3] and the Fuchs-Sondheimer[35,36] formula:

$$k(d \gg \Lambda_\omega, \lambda) = \frac{\frac{1}{3}\int d\omega C_\omega v_\omega \Lambda_\omega \frac{3}{\eta_\omega^2}\left\{1 - \frac{1}{\eta_\omega}\arctan(\eta_\omega)\right\}}{\frac{1}{C}\int d\omega C_\omega \frac{1}{\eta_\omega}\arctan(\eta_\omega)}$$

$$k(\lambda \gg \Lambda_\omega, d) = \frac{1}{3}\int d\omega C_\omega v_\omega \Lambda_\omega \left\{1 - \frac{3}{2}\mathrm{Kn}_\omega\left[\frac{1}{4} - E_3\left(\frac{1}{\mathrm{Kn}_\omega}\right) + E_5\left(\frac{1}{\mathrm{Kn}_\omega}\right)\right]\right\}$$

(17)

The suppression function is defined by the relation to the effective thermal conductivity[13,25] as $k = \frac{1}{3}\int C_\omega v_\omega \Lambda_\omega S_\omega d\omega$, where again the integral over $\omega$ implies a sum over all phonon polarizations and frequencies, in compact form[33]. The suppression function characterizes the effective reduction of the MFP of phonons due to size effects from the heat transfer geometry, which corresponds to a reduction in thermal conductivity. However, only in the large grating period limit can a universal suppression function, necessary for MFP reconstruction from the previously developed reconstruction algorithm[23], be obtained.

The temperature decay curves for some representative thicknesses of film and grating spacings for Si and PbSe are shown in Fig. 2, utilizing the same material property data from the previous one-dimensional TTG studies[3,13]. The variational approach is compared against the Monte Carlo results obtained for the thin film TTG[6]. The variational approach demonstrates a predictive ability to studying the temperature decay over a broad range of grating periods, even into the strongly non-diffusive thermal transport regime, when the thermal decay deviates from exponential decay.

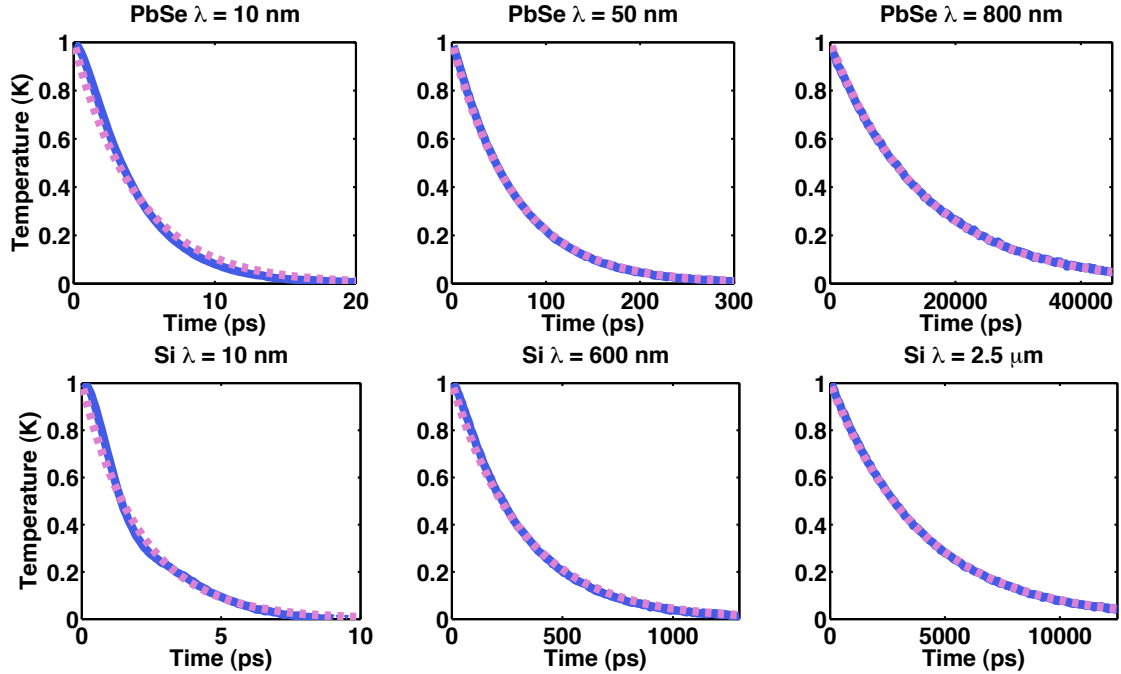

FIG 2: Temperature decay dynamics for PbSe (10 nm thick film) and Si (390 nm thick film) for several grating spacings, comparing the results of the variational technique (dashed line) and Monte Carlo runs (solid line).

To get a clearer picture of the onset of non-diffusive transport, we plot the thermal decay rate against the square of the grating wavevector in Fig. 3. As the slope of this relation is equal to the thermal diffusivity, in the diffusive regime the dependence is linear[6,11]. For larger grating wavevectors $q$ (i.e., at smaller periods $\lambda$ ), the solution visually deviates from the Fourier heat conduction model, and this deviation occurs at TTG periods on the order of 5 microns and 50 nanometers for Si and PbSe, at membrane thicknesses of 390 nm and 10 nm, respectively. The thermal decay rate gives a quantitative method to determine when the conductivity deviates from the bulk value for a given membrane thickness. Note that this thermal conductivity obtained for large grating periods will still experience size effects from the membrane boundary scattering as predicted by the Fuchs-Sondheimer model[35,36]. This is a different metric for determining the onset of non-diffusive transport than looking at the thermal conductivity accumulation function and looking at the value for MFP's below which contribute to 50% of the thermal conductivity[37]. While the thermal conductivity accumulation function would yield a simple estimate at what length scales one would expect to observe non-diffusive transport, the utilization of the decay rate is a more rigorous method of taking into account the full effect of the geometry and heat transfer configuration of the system.

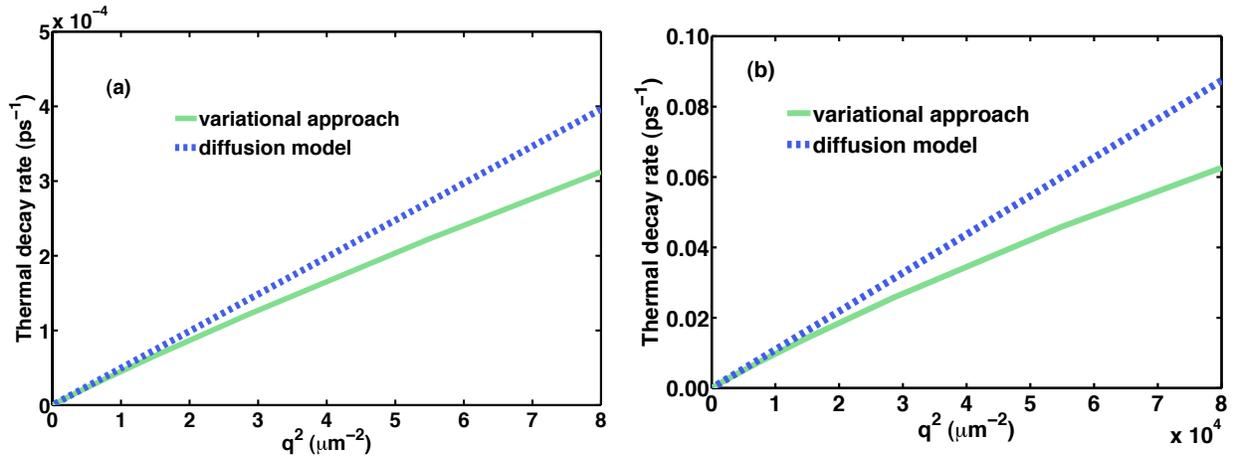

FIG. 3: Thermal decay rate plotted against wavevector squared for Si (a) and PbSe (b). The deviation from the Fourier conduction model on the order of roughly 10% occurs at squared wavevector values on the order of 2 and 20,000 $\mu m^{-2}$ for Si (390 nm membrane thickness) and PbSe (10 nm membrane thickness), respectively, which corresponds to on the order of 5 microns and 50 nm, respectively, to observe the onset of suppression of thermal conductivity.

The thermal decay rate from the Fourier heat conduction equation is directly related to the thermal conductivity, and utilizing this definition we extract the effective thermal conductivity as given by Eq. (16). By normalizing to the bulk thermal conductivity, we compare the effective normalized thermal conductivity as a function of the grating period for various thickness films. Fig. 4 shows the effective thermal conductivities of Si and PbSe for various thicknesses of film, plotted against the grating period. The material properties are the same as those utilized in our previous work[3] and Collins *et al.*[13]. There is excellent agreement with the Monte Carlo results across a broad range of film thicknesses and grating periods.

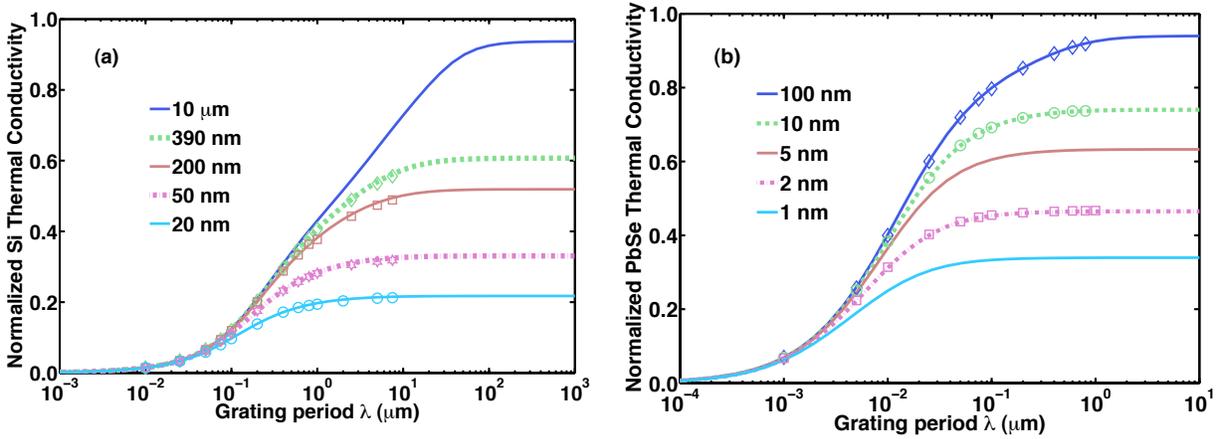

FIG 4: Normalized effective thermal conductivity as a function of the grating period for various thicknesses (increasing in thickness from bottom line to top line) of the thin film for (a) Si and (b) PbSe. The variational approach (lines) yields effective conductivities that show excellent agreement with Monte Carlo simulations (symbols) over a broad range of grating periods and film thicknesses.

We can determine the grating period yielding a 5% reduction in the effective thermal conductivity compared to the Fuchs-Sondheimer (long grating period) limit. This provides a quantitative metric of the onset of non-diffusive transport for a given membrane thickness. From Fig. 4(a), we see that for Si, the grating period that is at the onset of non-diffusive transport occurs at approximately 40 $\mu$m, 10 $\mu$m, 7 $\mu$m, 3 $\mu$m, 2 $\mu$m for membrane thickness of 10 $\mu$m, 390 nm, 200 nm, 50 nm, and 20 nm, respectively. Similarly for PbSe, Fig. 4(b) shows that the grating periods for which we can deduce the onset of non-diffusive transport occur at 300 nm, 120 nm, 90 nm, 60 nm, 50 nm for membrane thicknesses of 100 nm, 10 nm, 5 nm, 2 nm, and 1 nm, respectively. For both materials we see that the grating period for the onset of non-diffusive transport is shorter for thinner membranes, and this can be qualitatively explained by the reduction of the effective MFP of the systems due to the boundary scattering.

The variational approach demonstrates the simplicity of optimizing a trial solution from the diffusive temperature profile that the experimentalist fits to, rather than a brute force solution of the BTE followed by a fitting to a diffusive profile in order to extract material properties such as the effective thermal conductivity. This also provides the opportunity to obtain analytical solutions to the BTE for geometries that have never been obtained previously.

## IV. SUMMARY

We have demonstrated the ability of the variational method to solve a multi-dimensional, transient heat transfer configuration over a broad range of length scales. We solved the BTE to yield the spectral energy density as a functional of the temperature in the thin film TTG geometry. By utilizing a trial solution obtained from the Fourier heat conduction model with the effective thermal conductivity treated as a variational parameter, and imposing an energy conservation condition integrated over the membrane thickness and time, the effective thermal conductivity was obtained analytically. The variational method was shown to yield the expected limits in the large grating period case and large film thickness case, converging to the one-dimensional TTG result and Fuchs-Sondheimer formula, respectively. The variational approach demonstrates a simple and direct method for obtaining an approximate solution to the BTE resulting in an accurate, analytical formula for the temperature profile, thermal decay rate, and the corresponding effective thermal conductivity.

The results of this work further enhance the understanding of non-diffusive transport in the context of temperature decay within geometries comparable to the MFP of the materials studied. Furthermore, it demonstrates the ability of the variational method beyond simple one-dimensional geometries, and the variational approach can suitably be applied to other problems involving non-diffusive phonon-mediated heat transport in nanostructures. We provide a complete expression for the thermal conductivity that shows its full dependence on the material properties and the geometry of the system. The analytical form provided gives the ability to study a wide variety of materials efficiently in order to better understand non-diffusive transport.

Acknowledgment: The authors are thankful for helpful discussions with Dr. Jean-Philippe Peraud and Professor Nicolas Hadjiconstantinou. This material is based upon work supported as part of the "Solid State Solar-Thermal Energy Conversion Center (S3TEC)", an Energy Frontier Research Center funded by the U.S. Department of Energy, Office of Science, Office of Basic Energy Sciences under Award Number: DE-SC0001299/DE-FG02-09ER46577.